\documentclass[12pt]{iopart}
\usepackage[dvips]{graphics,graphicx}
\begin{document}

\title[Persistent current of atoms]{Persistent current of atoms
in a ring optical lattice}

\author{Andrey R. Kolovsky$^{1,2}$}

\address{$^1$Max-Planck-Institut f\"ur Physik komplexer Systeme, 01187 Dresden,
         Germany}
\address{$^2$Kirensky Institute of Physics, 660036 Krasnoyarsk, Russia}
\ead{kolovsky@mpipks-dresden.mpg.de}

\begin{abstract}
We consider a small ensemble of Bose atoms in a
ring optical lattice with weak disorder. The atoms are assumed
to be initially prepared in a superfluid state with non-zero
quasimomentum and, hence, may carry matter current.
It is found that the atomic current persists in time for a low value
of the quasimomentum but decays exponentially for a high
(around one quater of the Brillouin zone) quasimomentum.
The explanation is given in terms of low- and high-energy spectra
of the Bose-Hubbard model, which we describe using the Bogoliubov
and random matrix theories, respectively.
\end{abstract}

\maketitle

\section{Introduction}

Ultracold atoms in optical lattices constitute an intense research
activity both in experimental and theoretical physics.
Up to now this system has mostly been used for modelling the
fundamental Hamiltonians of solid state theory (see, \cite{Grei02,Jaks05},
for example) where the number of particles is macroscopically large.
However, the recent progress with manipulating a countable
number of atoms \cite{Kuhr01,Chuu05} makes it possible
to build a system of arbitrary size, ranging from microscopic
to macroscopic. In this border region between microscopic
and macroscopic one has to deal with a finite number of
atoms which, on the one hand, is too large to use
the single-particle approach but, on the other hand,
is too small to justify the thermodynamic limit. In the
present work we theoretically analyse one of these problem
related to superfluidity of a few ($N\sim10$) Bose atoms
in a ring optical lattice \cite{Amic05} with a few ($L\sim 10$) sites.

It should be stressed in the very beginning that, currently,
there are two different definitions of superfluidity in the
physics literature. One definition is based on the system's responce
to a phase twist. With respect to Bose atoms in
a lattice this approach is discussed, in particular, in Ref.~\cite{Roth04},
and a method of how one can realize the twisted boundary conditions in 
a laboratory experiment is suggested in Ref.~\cite{Amic05}. The other 
definition originates in the Landau criterion of superfluidity
and involves a responce of a superfluid flow to `wall roughness' 
\cite{Land41,Astr04}. In this work we try to reconcile both
approaches. Specifically, we address the following problem.
Assume that we have $N$ Bose atoms in a ring lattice with $L$ 
sites in a superfluid state with given quasimomentum $\kappa=2\pi k/L$:
\begin{equation}
\label{2}
|\kappa\rangle=\left(\frac{1}{\sqrt{L}}\sum_l \hat{a}^\dag_l
e^{i\kappa l}\right)^N |0\rangle \;.
\end{equation}
We are interested in the time evolution of this state
(which we also shall refer to as the supercurrent state)
in the presence of a weak scattering potential and atom-atom interactions.

We note that for a BEC of atoms ($N\gg 1$)
the problem of superfluid atomic current has been considered
in a large number of papers (see
Refs.~\cite{Wu03,Smer02,Kono02,Fall04,Cata03,Scot04,preprint,Polk05},
to cite few of them). The starting point of all these studies
is the mean-field approach, which is sometimes rectified by taking into
account the quantum fluctuations \cite{preprint,Polk05}.
The mean-field theory predicts a destruction of the supercurrent
as soon as the quasimomentum exceeds one quater of the reciprocal
lattice constant ($\kappa>\pi/2$ in the notations used).
In order to justify the mean-field approach in a 1D lattice the
mean number of atoms per one site should be much larger than unity.
As stated above, in the present work we focus on the opposite
limit $N/L\sim1$, where the mean-field approach is not applicable.
For this reason we treat cold atoms in an optical lattice
from a different viewpoint, in a sense closer to quantum
optics than to condensed matter physics.

The paper essentially consists of two parts, -- in the first part
(Sec.~\ref{sec2}), after a brief preliminary analysis, 
we report the results of numerical simulations of the
system dynamics, and in the second part (Sec.~\ref{sec3} and Sec.~\ref{sec4})
we explain the observed regimes in terms of the energy spectrum of 
the system. The main  results are  summarized in the concluding 
Sec.~\ref{sec5}.

\section{Supercurrent dynamics}
\label{sec2}
Before proceeding with numerical simulations, we shall briefly
discuss possible regimes for the atomic current.

\subsection{Preliminary analysis}
Let us first consider the single-particle problem. In the tight-binding
approximation the Hamiltonian of the system reads
\begin{equation}
\label{3}
\widehat{H}=-\frac{J}{2}\sum_l (|l+1\rangle\langle l|+h.c.)
+\sum_l V_l|l\rangle\langle l| \;,
\end{equation}
where $|l\rangle$ are the Wannier functions, $J$ the hopping matrix
element, and $V_l$ the random scattering potential. In what follows,
to be concrete, we shall consider $0\le V_l\le \epsilon$ with $\epsilon\ll J$.
The operator $\widehat{V}=\sum_l V_l|l\rangle\langle l|$ couples the degenerate
states with opposite quasimomentum, resulting in new eigenstates
$|\kappa_{c,s}\rangle=(|\kappa\rangle \pm |-\kappa\rangle)/\sqrt{2}$
with an energy splitting
$|E_c-E_s|=2|\langle\kappa|\widehat{V}|-\kappa\rangle|=2|V(2k)|$, where
\begin{displaymath}
V(k)=\frac{1}{L}\sum_l V_l\exp\left(\frac{2\pi k}{L}l\right)
\sim\frac{\epsilon}{L}
\end{displaymath}
is the Fourier transform of $V_l$.
Thus, in course of time, an atom in a ring will periodically change its
momentum to the opposite one with the frequency $\Omega_\epsilon\sim 
\epsilon/\hbar L$. It is worth of stressing that this periodic
dynamics is exclusively due to the finiteness of $L$ and the assumed
condition $\epsilon\ll J$, which means that $\widehat{V}$
couples only the degenerate states of the unperturbed system. 
\footnote{In terms of Anderson's localization theory the
above conditions mean that the Anderson localization length
is much larger than the system size.}

Next we consider the multi-particle case,
\begin{equation}
\label{4}
\widehat{H}=
-\frac{J}{2}\sum_l \left( \hat{a}^\dag_{l+1}\hat{a}_l+h.c.\right)
  +\frac{U}{2}\sum_l \hat{n}_l(\hat{n_l}-1)
  +\sum_l V_l\hat{n}_l \;,
\end{equation}
where $\hat{a}^\dag_{l}$ and $\hat{a}_l$ are the bosonic creation
and annihilation operator, $\hat{n}_l=\hat{a}^\dag_{l}\hat{a}_l$,
and $U$ is the on-site interaction energy.
Using the canonical transformation,
$\hat{b}_k=(1/\sqrt{L})\sum_l \exp(i2\pi kl/L)\hat{a}_l$, it is
convenient to present the Hamiltonian (\ref{4}) in the form
\begin{eqnarray}
\label{5}
\widehat{H}=-J\sum_k \cos\left(\frac{2\pi k}{L}\right)
\hat{b}_k^\dag\hat{b}_k +\frac{U}{2L}\sum_{k_1,k_2,k_3,k_4}
\hat{b}_{k_1}^\dag\hat{b}^\dag_{k_2}\hat{b}_{k_3}\hat{b}_{k_4}
\tilde{\delta}(k_1+k_2-k_3-k_4)\nonumber\\
+\sum_{k_1,k_2}V(k_1-k_2)\hat{b}^\dag_{k_1}\hat{b}_{k_2} \;,
\end{eqnarray}
where $\tilde{\delta}(k)=1$ if $k$ is a multiple of $L$, and
$\tilde{\delta}(k)=0$ otherwise. For $U=0$ and $\epsilon=0$ 
the multi-particle eigenstates of the system (\ref{5}) are
the quasimomentum Fock states $|{\bf n}\rangle=|n_0,n_1,\ldots,n_{L-1}\rangle$,
where $\sum_k n_k=N$. Our state of interest corresponds to
$|\kappa\rangle=|\ldots,0,N_k,0,\ldots\rangle$, where all atoms
have one and the same quasimomentum. Similar to the single-particle
case, the random potential couples 
this state to the supercurrent state with the opposite quasimomentum
$|-\kappa\rangle=|\ldots,0,N_{k'},0,\ldots\rangle$, $k'={\rm mod}_L(-k)$.
However, now the coupling is indirect and involves the intermidiate
states $|\kappa(m)\rangle=|\ldots,(N-m)_k,\ldots,m_{k'},\ldots\rangle$,
as it immediately follows from the explicit form of the scattering potential 
in the momentum representation. Thus the time evolution
of the state $|\kappa\rangle$ is defined by the following $(N+1)\times(N+1)$
matrix,
\begin{equation}
\label{6}
A_{m,m'}=E_m\delta_{m,m'}+\sqrt{(N-m)(m+1)}
\left[V(2k)\delta_{m+1,m'}+V^*(2k)\delta_{m,m'+1}\right] \;,
\end{equation}
where $E_m=E_\kappa\equiv-JN\cos\kappa$ are the degenerate energies of
the states $|\kappa(m)\rangle$ and the next terms the transition
matrix elements $\langle\kappa(m)|\widehat{V}|\kappa(m')\rangle$.
The spectrum of the matrix (\ref{6}) is equidistant with
the level spacing $2|V(2k)|$. Thus we have reproduced the result of the
single-particle analysis, where the time evolution of the system is periodic with
the frequency $\Omega_\epsilon\sim \epsilon/\hbar L$.

Now we switch on the interaction. Then the intermidiate
states $|\kappa(m)\rangle$ aquire energy shifts $E_m=E_m(U)$, which appear
to be $m$-dependent. Using the first order perturbation theory we obtain
\begin{eqnarray}
E_m=E_\kappa + \frac{U}{2L}\langle\kappa(m)|
\sum \hat{b}_{k_1}^\dag\hat{b}^\dag_{k_2}\hat{b}_{k_3}\hat{b}_{k_4}
\tilde{\delta}(k_1+k_2-k_3-k_4)|\kappa(m)\rangle \nonumber \\
=E_\kappa + \frac{U}{2L}\left[(N-m)(N-m-1)+m(m-1)+4(N-m)m\right] \nonumber \\
\approx E_\kappa +\frac{UN^2}{2L}+\frac{U}{L}m(N-m) \;.
\label{7}
\end{eqnarray}
Due to the mismatch of the energy levels $E_m$, the supercurrent
states $|\kappa\rangle\equiv|\kappa(0)\rangle$ and
$|-\kappa\rangle\equiv|\kappa(N)\rangle$ become effectively decoupled and,
hence, the supercurrent should persist in time.

At this point we would like to note the analogy of the
problem discussed with that for a BEC in double well potential
\cite{doublewell}. Drawing this analogy further
we can estimate the minimal $U_{min}$ required for stabilization
of the suppercurrent as
\begin{equation}
\label{9}
U_{min}\approx 8\epsilon/N \;.
\end{equation}
For the sake of completeness we present a derivation of
the estimate (\ref{9}) in the next subsection (which can be safely
skipped if a reader is familiar with the subject).

\subsection{Semiclassical approach}
\begin{figure}[t]
\center
\includegraphics[height=8.5cm, clip]{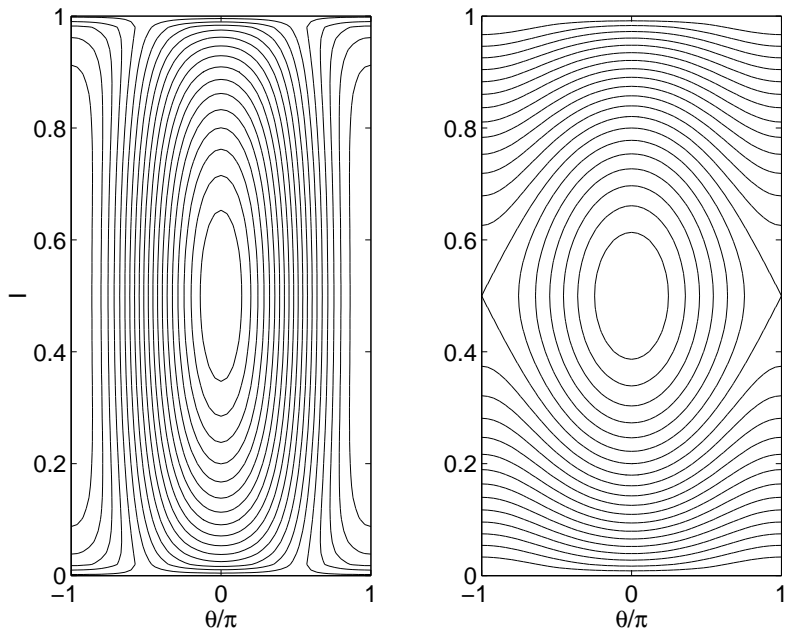}
\caption{Phase portrait of the effective system (\ref{8}) for
$g/|V|=1$ (left panel) and $g/|V|=10$ (right panel).}
\label{fig5b}
\end{figure}

The standart method of treating the system (\ref{6}-\ref{7})
consist of mapping it onto an effective classical system
(terms proportional to the identity matrix are omitted),
\begin{equation}
\label{8}
H_{eff}=gI(1-I)+2|V|\sqrt{I(1-I)}\cos\theta \;,\quad g=UN/L \;,
\end{equation}
followed by a semiclassical quantization, where $1/N$ plays
the role of Planck's constant. The phase portrait
of the system (\ref{8}) is shown in Fig.~\ref{fig5b} for
$g/|V|=1$ and $g/|V|=10$. It is seen that when $g$ is increased
the phase portrait becomes similar to that of
classical pendulum, with a separatrix separating the librational
and rotational regimes. The maximal and minimal values of the
classical action $I$ along the separatrix are given by
$I^*\approx 1/2\pm\sqrt{|V|/2g}$. The quantum states, associated with
$I$, are decoupled only if they lie above the separatrix.
Then, by requiring $|I^*-1/2|\le 1/2$ and noting that 
$|V|\sim \epsilon/L$, we come to the estimate (\ref{9}).

\begin{figure}[t]
\center
\includegraphics[height=8.5cm, clip]{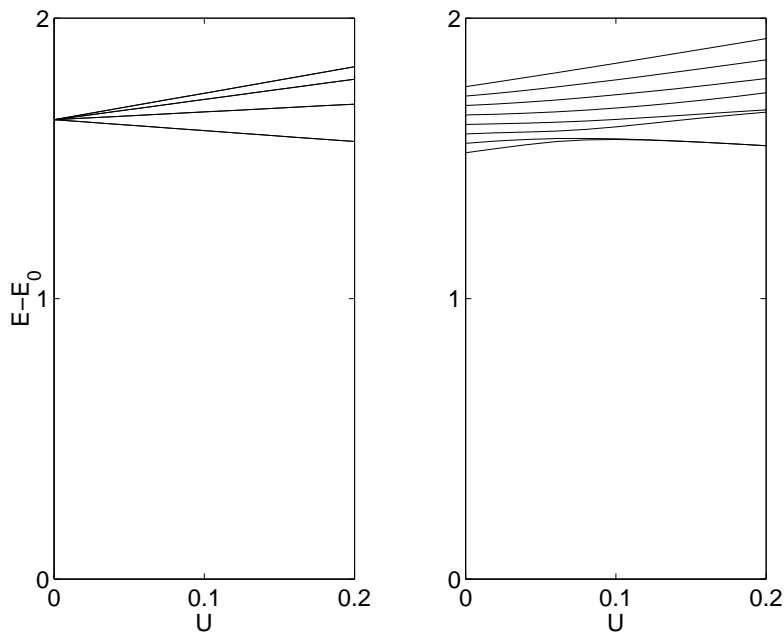}
\caption{Spectrum of the matrix (\ref{6}-\ref{7}) as function
of the interaction constant $U$. (The energy is measured in
units of $J$, $E_0=-JN+UN^2/2L$.) Parameters are $N=7$, $L=9$, $\kappa=2\pi/L$,
and $|V(2)|=0$ (left panel) and $|V(2)|=0.0168$ (right panel).}
\label{fig7b}
\end{figure}

Needless to say, the semiclassical approach described above
requires $1/N\ll 1$ and is not accurate for small $N$.
Nevetheless, even for $N\sim10$ the spectrum of the matrix
$A$ can be well understood in terms of the effective system (\ref{8}).
For the purpose of future reference, the right panel in Fig.~\ref{fig7b}
shows the numerical solution of the matrix eigenvalue problem for
$N=7$, $L=9$, and $|V|=0.0168$.
In particular, at $U=0.2J$ one can identify
the first four top levels with the phase trajectories
below the separatrix, next two levels with trajectories
around the separatrix, and the last two almost degenerate
levels with trajectories well above the separatrix.

\subsection{Numerical results}
The above conclusion about the persistent current relies on the 
applicability of a perturbative approach. Formally
this means that the supercurrent state $|\kappa\rangle$, as well
as the intermidiate states $|\kappa(m)\rangle$, have to be
approximate eigenstates of the system at $\epsilon=0$. This imposes
the upper boundary $U_{max}$ on the interaction constant, which appears 
to depend on the quasimomentum $\kappa$. Indeed, the state $|\kappa\rangle$ with
the energy $E_\kappa\approx-JN\cos\kappa$ is an approximate eigenstate of the
system only if $U$ is smaller than the characteristic energy gap separating 
it from the other energy states, coupled to $|\kappa\rangle$ by interaction.
As the first guess one can set this gap to the mean level spacing, given
by the inverse density of state $\overline{\Delta E}= 1/f(E)$. It is
easy to show that for $U/J\le 1$ the density of states of (\ref{4})
is given by the Gaussian distribution (see Fig.~\ref{fig2} below)
\begin{equation}
\label{10}
f(E)\approx\frac{{\cal N}}{\sqrt{2\pi}\sigma}
\exp\left[\frac{(E-\bar{E})^2}{2\sigma^2}\right] \;,
\end{equation}
where ${\cal N}=(N+M-1)!/N!(M-1)!$ is the dimension of the Hilbert space,
$\sigma\sim J\sqrt{N}$ and ${\bar E}\sim UN^2/L+\epsilon N/2$.
Thus the characteristic gap for a supercurrent state, which belongs to the
central part of the spectrum (i.e., for $\kappa\sim \pi/2$),
is essentially smaller than that for a supercurrent state with
low quasimomentum $\kappa\ll \pi/2$. As a consequence, $U_{max}$
for the supercurrent state with $\kappa\sim \pi/2$ may be smaller than $U_{min}$.
In the other words, the perturbative approach  of Sec.~2.1
(where we used first order perturbation theory to find
corrections to the eigenenergies of states $|\kappa(m)\rangle$)
breaks down before the stabilization of the supercurrent is achieved.
\begin{figure}[t]
\center
\includegraphics[height=8.5cm, clip]{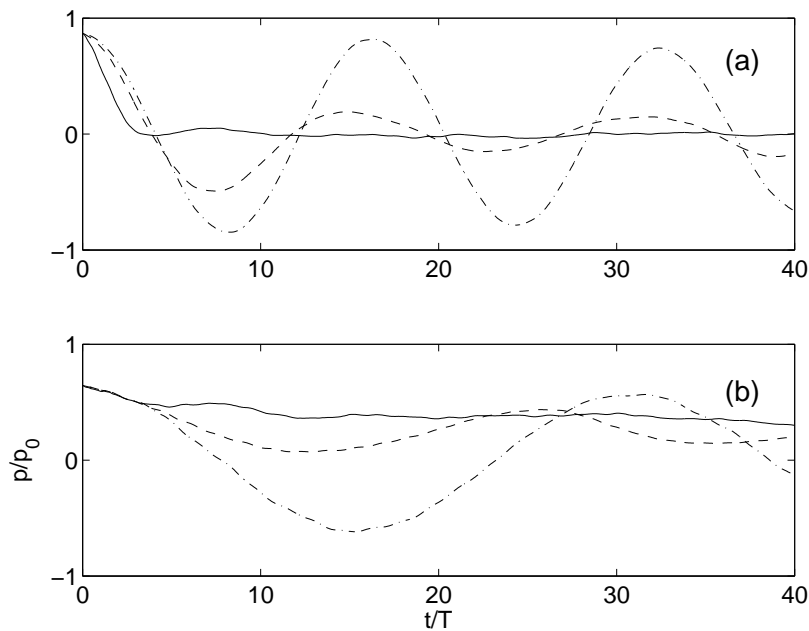}
\caption{The mean momentum of $N=7$ atoms in a lattice with $L=9$
sites. The magnitude of the scattering potential $\epsilon=0.2$. The
interaction constant $U=0.02$ (dash-dotted lines), $U=0.1$ (dashed lines),
and $U=0.2$ (solid lines). The atoms are initially
prepared in the supercurrent state (\ref{2}) with $\kappa=6\pi/L$ (upper
panel) and $\kappa=2\pi/L$ (lower panel).}
\label{fig1}
\end{figure}

Figure \ref{fig1} compares the dynamics of $N=7$ atoms in
a lattice with $L=9$ sites, which were initially prepared in the supercurrent
state (\ref{2}) with high, $\kappa=6\pi/L$, and low, $\kappa=2\pi/L$,
quasimomentum. The normalized mean momentum of the atoms, $p(t)=N^{-1}
{\rm Im}[\langle\Psi(t)|\sum_l \hat{a}^\dag_{l+1}\hat{a}_l|\Psi(t)\rangle]$,
is depicted. It is seen in the upper panel of Fig.~\ref{fig1} that
in the former case of high quasimomentum the oscillatory behaviour
of $p(t)$ changes to irreversible decay as the interaction constant
is varied from $U=0.02$ to $U=0.2$. (From now on we set $J=1$, i.e.,
energy is measured in the units of $J$ and time in the units
$T=2\pi\hbar/J$.) Further increase of the
interaction constant (results are not shown) is  reflected in even
faster decay of $p(t)$. This should be contrasted with the case of low
quasimomentum (lower panel), where the current oscillations at $U=0.02$
change to persistent current at $U=0.2$. Here further increase of
$U$ leaves the system dynamics qualitatively unchanged at least
till $U=1$. The displayed numerical results suggest that the perturbative
approach of Sec.~2.1 works for $\kappa=2\pi/L$ but does not
work for $\kappa>2\pi/L$. We shall come back to this point later
on in Sec.~4.3.

It is worth stressing that through the paper we consider
a single realization for the random potential (i.e., no average
over disorder). Specifically to the considered lattice of
$L=9$ sites, the random entries are
$V_l=\epsilon (0.80,0.59,0.06,0.18,0.97,0.31,0.67,0.78,0.49)$.
The Fourier transform of this sequence gives
$|V(2k)|=0.084\epsilon$ and $|V(2k)|=0.144\epsilon$ for $k=1$
and $k=3$, respectively.

\section{High-energy spectrum}
\label{sec3}
To get a better insight in the physics of the discussed phenomena
we shall discuss the displayed in Fig.~\ref{fig1} results in
terms of the energy spectrum of the system (\ref{4}). We begin
with the case of a high quasimomentum which, as mentioned above,
refers to the central part of the spectrum.

\subsection{Spectral statistics}
We have found that in the case of high initial quasimomentum
a transition from oscillatory dynamics to irreversible decay
is associated with the transition to chaos in the Bose-Hubbard model. 
Following Ref.~\cite{66}, we
shall monitor this transition by analysing the distribution
of distances between the neighbouring levels, normalised to the
mean level spacing: $s=(E_{n+1}-E_n)/\overline{\Delta E}=
(E_{n+1}-E_n)f[(E_{n+1}+E_n)/2]$. It should be stressed that
the presence of random potential in the Hamiltonian (\ref{4}) alone 
does not yet induced chaos in the system. The only consequence of 
a weak disorder (relevant to the spectral statistics) is that
it breaks the translational symmetry and, hence,
we need not worry about decomposition of the energy spectrum
into the independent subsets (labeled, in the absence of a
random potential, by total quasimomentum of the atoms \cite{66}).

\begin{figure}[t]
\center
\includegraphics[height=8.5cm, clip]{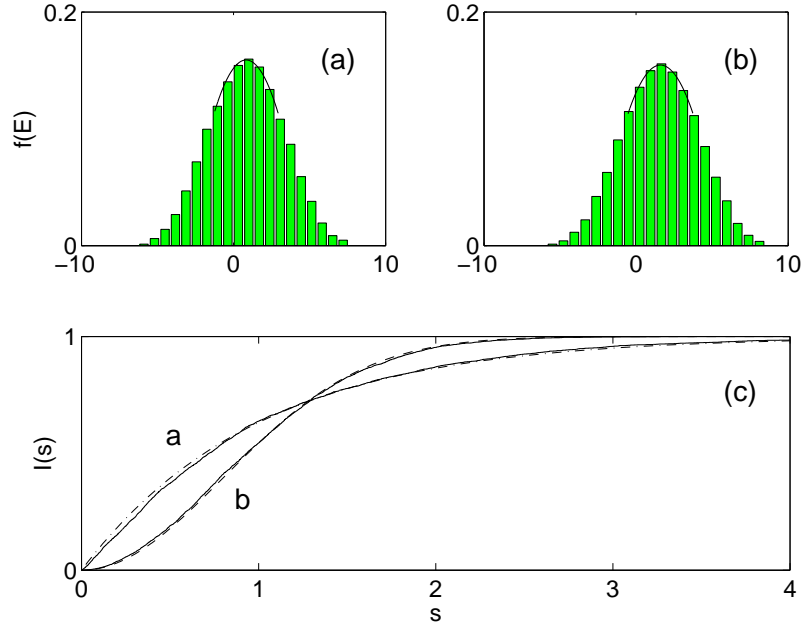}
\caption{(a,b) -- density of states of $N=7$ atoms in a lattice with $L=9$
sites for the interaction constant $U=0.02$ and $U=0.2$, respectively.
The magnitude of the scattering potential $\epsilon=0.2$.
(c) -- integrated level spacing distributions for the central part
of the spectrum.}
\label{fig2}
\end{figure}

The results of the statistical analysis of the high-energy spectrum 
are presented in Fig.~\ref{fig2}. The dash-dotted and dashed lines in 
panel (c) correspond to the integrated distribution, $I(s)=\int_0^s P(s')ds'$,
for the Poisson statistics,
\begin{equation}
\label{11}
P(s)=\exp(-s) \;,
\end{equation}
which is typical for a generic integrable system, 
and the Wigner-Dyson statistics,
\begin{equation}
\label{12}
 P(s)=\frac{\pi}{2}s\exp\left(-\frac{\pi}{4}s^2\right) \;,
\end{equation}
typical for non-integrable systems. These distributions reflect
the different character of the parametric dependence of the
energy levels $E_n=E_n(\lambda)$ on some parameter in the
Hamiltonian ($\lambda=U$ in our case).
Namely, in the integrable case the energy levels may cross and,
hence, one finds an arbitrary small $s$. On the contrary,
if the system is non-integrable, the energy levels show avoided
crossings and probability of finding small $s$ tends to zero.

The panel (a) in Fig.~\ref{fig2} shows the density
of states $f(E)$ for $U=0.02$, where only the data from the central
part of the spectrum (marked by the inverse parabola) were
used for the statistical analysis. It is seen in the lower panel that for 
$U=0.02$ the level spacing distribution follows the Poisson statistics.
Thus for this value of the interaction constant the system should be
classified as integrable, which is consistent with the periodic
dynamics of the mean momentum in Fig.~\ref{fig1}(a).
The panel (b) in Fig.~\ref{fig2} shows the density of states
for $U=0.2$. Apart from an uniform shift of the spectrum to positive
values, no qualitative change in $f(E)$ is observed.
However, we do observe a qualitative change in the level spacing
distribution. Now it reliably follows the Wigner-Dyson statistics 
which, as mentioned above, is a hallmark of quantum chaos.

\subsection{Local density of states}
The spectral statistics is only one (and, in fact, rather poor)
characteristic of the system. In particular, the level spacing distribution 
remains unchanged (Wigner-Dyson) in the interval $0.2\le U\le 1$, although
the decay rate of the supercurrent changes with $U$. One gets more
information about the system  by studying its eigenfunctions. To this
end we introduce a quantity $R(m,n)$,
\begin{equation}
\label{13}
R(m,n)=|\langle \Psi_m(U')|\Psi_n(U)\rangle|^2 \;,
\end{equation}
closely related to the so-called local density of states. 
\footnote{The local density of states is defined as
$R(m,E)=\sum_n R(m,n)\delta(E-E_n)$.}
In Eq.~(\ref{13}) $|\Psi_n(U)\rangle$ are the eigenfunctions
of the Hamiltonian (\ref{4}) calculated for a given $U$ and ordered according
to their energies. In what follows we shall fix $U'=0.02$ while $U$ will be scanned
in the interval $0.2\le U\le 1$. Since for $U=0.02$ the system
is integrable, the matrix (\ref{13}) can be alternatively viewed
as the matrix of the expansion coefficients of the chaotic states
$|\Psi_n(U)\rangle$ over `regular basis' $|m\rangle=|\Psi_m(U=0.02)\rangle$.

\begin{figure}[t]
\center
\includegraphics[height=8.5cm, clip]{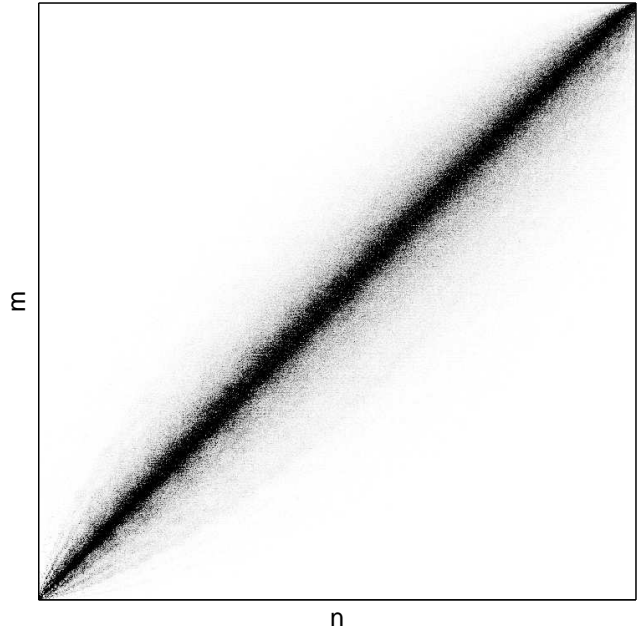}
\caption{Gray-scale image of the matrix (\ref{13}).
(The system parameters are the same as in Fig.~\ref{fig1} and Fig.~\ref{fig2}.)}
\label{fig3}
\end{figure}

\begin{figure}[t]
\center
\includegraphics[height=8.5cm, clip]{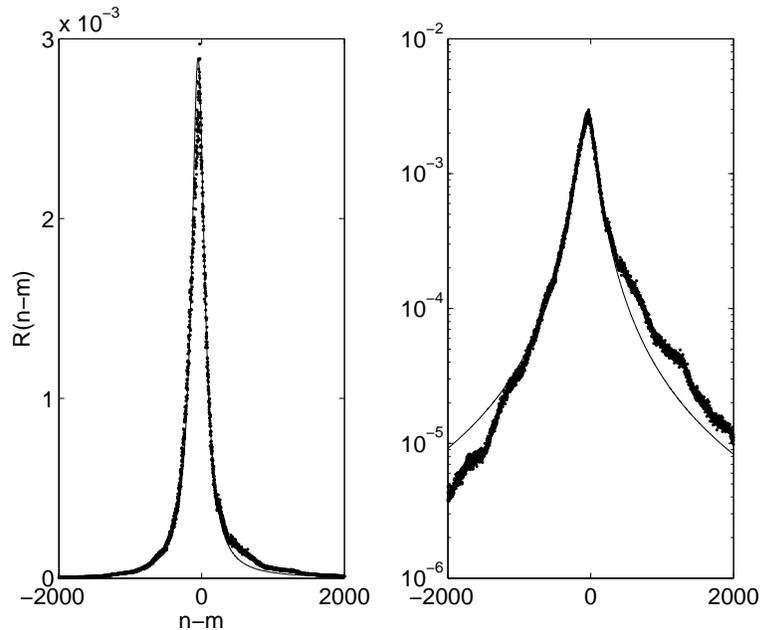}
\caption{Mean values of the matrix elements across
the main diagonal in the central part of the matrix. The solid
line is the best fit by the Breit-Wigner formula (\ref{15}).}
\label{fig4}
\end{figure}

The characteristic structure of the matrix (\ref{13}) is shown in 
Fig.~\ref{fig3} for $U=0.2$. It is seen that that $R$ is a banded matrix
with strongly fluctuating matrix elements. The mean values of the elements 
across the main diagonal,
\begin{equation}
\label{14}
\bar{R}(\Delta m)=\frac{1}{M}\sum_{m=-M/2}^{M/2} R(m,m+\Delta m) \;,
\quad \sum_{\Delta m} \bar{R}(\Delta m)=1 \;,
\end{equation}
are shown in Fig.~\ref{fig4} on linear and logarithmic scales.
(Here, as in the spectrum analysis, we consider an energy window
of the order of unity in the central part of the spectrum.) It is seen that 
$\bar{R}(\Delta m)$ converges to the Lorentzian, 
\footnote{ The distribution (\ref{15}) is typical for the banded random 
matrices \protect\cite{Fyod96}. It is interesting to note in this connection
that for the $N/L\sim 1$ neither matrix of
the Hamiltonian (\ref{4}) nor that of the Hamiltonian (\ref{5})
are banded. It is an open problem in the random matrix theory
to extend the results of \protect\cite{Fyod96} to the present case
of very sparse but not banded matrices.}
%
\begin{equation}
\label{15}
\bar{R}(\Delta m)=\frac{\Gamma/2\pi}{(\Delta m)^2+\Gamma^2/4} \;.
\end{equation}
We note, in passing, that a similar result is reported in
the recent paper \cite{Hill06} devoted to the spectral properties
of the three-site Bose-Hubbard model.

The distribution (\ref{15}), also known
as the Breit-Wigner formula, implies the exponential decay
of the supercurrent state. Indeed, considering the overlap
integral $\langle\kappa|\kappa(t)\rangle$, one has
\begin{eqnarray}
\fl \langle\kappa|\exp\left(-\frac{i}{\hbar}\widehat{H}t\right)|\kappa\rangle
=\sum_{m,m',n} \langle\kappa|m\rangle\langle m|\Psi_n\rangle
\exp\left(-\frac{i}{\hbar}E_n t\right)
\langle \Psi_n|m'\rangle\langle m'|\kappa\rangle \nonumber \\
\approx \frac{1}{N+1}\sum_m R(m,n)\exp\left(-\frac{i}{\hbar}E_n t\right)
\sim \sum_{\Delta m} \bar{R}(\Delta m)
\exp\left(-\frac{i\overline{\Delta E}t}{\hbar}\Delta m \right)
\nonumber \;,
\end{eqnarray}
where we substitute the exact energy levels $E_n$ by their approximate
positions, $E_n\approx E_\kappa+\overline{\Delta E}\Delta m$.
(This approximation obviously holds till time
$t_\hbar\sim\hbar/\overline{\Delta E}=\hbar f(E_k)$, which
increases exponentially with the system size.) Substituting here
$\bar{R}(\Delta m)$ from Eq.~(\ref{15}) we have $\langle\kappa|\kappa(t)\rangle
=\exp(-\Gamma \overline{\Delta E} t/\hbar)$.
We found that the width $\Gamma$ grows approximately quadratically
with $U$ in the interval $0.2\le U\le 1$.

\section{Low-energy spectrum}
\label{sec4}

We turn to the case of low quasimomentum. For small
$\kappa$ the energy of the supercurrent state (\ref{2})
falls into the low-energy tail of the density of states (\ref{10}),
where the random matrix approach is not applicable.
On the other hand, the low-energy spectrum of the interacting
Bose atoms is believed to be described by the Bogoliubov theory.
For this reason we review the Bogoliubov approach for
a finite size system. Through the section, if not
stated otherwise, we assume the homogeneous case $\epsilon=0$.

\subsection{Bogoliubov approach}
As an intermidiate step, let us show that the Bogoliubov approach
amounts to the following two assumptions. (i) The low energy
eigenstates of interacting Bose atoms are given by a linear
superposition of the quasimomentum Fock states, where $n$ atoms
have quasimomentum $\kappa$, $n$ quasimomentum $-\kappa$, and the
rest $N-2n$ have zero quasimomentum \cite{Legg01}, i.e.,
\begin{equation}
\label{21}
|\Psi_\kappa\rangle=\sum_{n=0}^{N/2} c_n
|N-2n,\ldots,n_k,\ldots,n_k',\ldots,0\rangle \;,\quad
k'={\rm mod}_L(-k) \;.
\end{equation}
(ii) The number of atoms with $\kappa\ne0$ is small compared to
the number of atoms with zero quasimomentum, i.e.,  only the coefficients
$c_n$ with $n\ll N/2$ have non-negligible values. (This condition is
automatically satisfied if one assumes the thermodynamic limit 
$N\rightarrow\infty$, $U\rightarrow0$, $UN=const$.)

The analysis goes as follows.
Substituting the wave function (\ref{21}) in the eigenvalue equation
with the Hamiltonian (\ref{5}),  we get a system of linear equations
for the coefficients $c_n$,
\begin{equation}
\label{22}
(2E_k n +\frac{U}{L} a_n)c_n
+\frac{U}{L}(b_n c_{n-1}+b_{n+1}c_{n+1})=E c_n \;,
\end{equation}
where
\begin{eqnarray}
a_n=2nN-3n^2+n+N(N-1)/2\approx n(2N-3n)+N^2/2 \;,\nonumber\\
b_n=(n+1)\sqrt{(N-2n)(N-2n-1)}\approx n(N-2n) \;,\nonumber
\end{eqnarray}
and $E_k=J[1-\cos(2\pi k/L)]$ is the single-particle excitation energy 
(should not be missmatched with the energy of the supercurrent
state, $E_\kappa=-JN\cos\kappa$). Assuming the thermodynamic limit,
Eq.~(\ref{22}) simplifies to
\begin{equation}
\label{23}
2(E_k + g)n c_n + gn c_{n-1}+g(n+1) c_{n+1} = E c_n \;,\quad
g=NU/L \;.
\end{equation}
Next, introducing the generating function,
\begin{displaymath}
\Phi(\theta)=\frac{1}{\sqrt{2\pi}}\sum_{n=-\infty}^\infty c_n e^{in\theta} \;,
\end{displaymath}
we present the system of linear equations (\ref{23}) as
a differential equation on the function $\Phi(\theta)$, 
\footnote{To be regorous, Eq.~(\ref{24}) is not strictly equivalent
to Eq.~(\ref{23}) in the sense that it also has solutions
with negative $E$.}
%
\begin{equation}
\label{24}
g\left[\hat{n}e^{i\theta}+2(1+\epsilon)\hat{n}+e^{-i\theta}\hat{n}\right]
\Phi(\theta)=E\Phi(\theta) \;,
\end{equation}
where $\hat{n}=-i\partial/\partial\theta$ and
$\varepsilon=E_k/g$. The general solution of (\ref{24}) reads
\begin{equation}
\label{25}
\Phi(\theta)=C\exp\left(i\int_0^\theta\frac{E/g-e^{i\vartheta}}
{2\cos\vartheta+2+\varepsilon} d\vartheta \right) \;.
\end{equation}
Finally, requiring $\Phi(\theta+2\pi)=\Phi(\theta)$ and calculating
the relevant integral,
\begin{displaymath}
\frac{1}{2\pi}\int_0^{2\pi}\frac{d\theta}{2\cos\theta+2+\varepsilon}
= \frac{1}{2\sqrt{(1+\varepsilon)^2-1}} \;,
\end{displaymath}
we get the equidistant spectrum with the transition frequency
\begin{equation}
\label{26}
\omega_k=2\sqrt{2gE_k+E_k^2} \;.
\end{equation}
The result (\ref{26}) reproduces the famous Bogoliubov equation
for the quasiparticle excitations of the Bogoliubov vacuum.

\subsection{Bogoliubov spectrum}
In the previous subsection we have considered an
excitation of the given quasimomentum state, with
the single-particle excitation energy $E_k=J(1-\cos\kappa)$.
To include the other quasimomentum states, the ansatz (\ref{21})
should be generalized to
\begin{equation}
\label{27}
|\Psi\rangle=\sum_{\bf n} c_{\bf n}
|N-2\sum_k n_k,n_1,n_2,\ldots\rangle \;,
\end{equation}
where ${\bf n}=(n_1,\ldots,n_{L/2})$.  Substituting (\ref{27})
in the stationary Schr\"odinger equation with the Hamiltonian
(\ref{5}), we obtain a system of rather complex equations on
the coefficients $c_{\bf n}$,
which can be solved analytically only in the thermodynamic limit.
In this limit, as it is easy to show, the whole
eigenvalue problem factorizes to $L/2$ eigenvalue
problems of the form (\ref{23}) and, hence, the whole spectrum 
is given by the direct sum of $L/2$ linear spectra.

A remark about the total quasimomentum, which is a global symmetry
of the system in the absence of random potential, is in turn. 
The substitution (\ref{21}) corresponds to zero total
quasimomentum. To get non-zero values of the total quasimomentum,
one should use a slightly different ansatz,
\begin{eqnarray}
\label{28}
|\Psi_\kappa\rangle=\sum_{n=0}^{N/2} c_n
|N-2n-m,\ldots,n_k+m,\ldots,n_k',\ldots,0\rangle \;,\quad \\
k'={\rm mod}_L(-k) \;,\quad m=1,\ldots,L-1 \;. \nonumber
\end{eqnarray}
Ansatz (\ref{28}) leads to the eigenvalue equation of the form (\ref{22})
but with different coefficients $a_n$ and $b_n$. 
In particular, considering the thermodynamic limit, Eq.~(\ref{23})
changes to
\begin{eqnarray}
\label{29}
\fl 2(E_k + g)(n+m) c_n + g\sqrt{n(n+m)}c_{n-1}+g\sqrt{(n+1)(n+m+1)}
c_{n+1} = E c_n \;.
\end{eqnarray}
We note, in passing, that if the spectra associated with different single-particle
excitation energy and different total quasimomentum are superimposed,
\begin{eqnarray}
\label{29a}
E=\sum_{k,m} \left\{ E^{(k,m)}(g)\right\} \;,
\end{eqnarray}
one typically finds a multiple degeneracy of the levels at $U=0$
(see Fig.~\ref{fig7} below).
\begin{figure}[t]
\center
\includegraphics[height=8.5cm, clip]{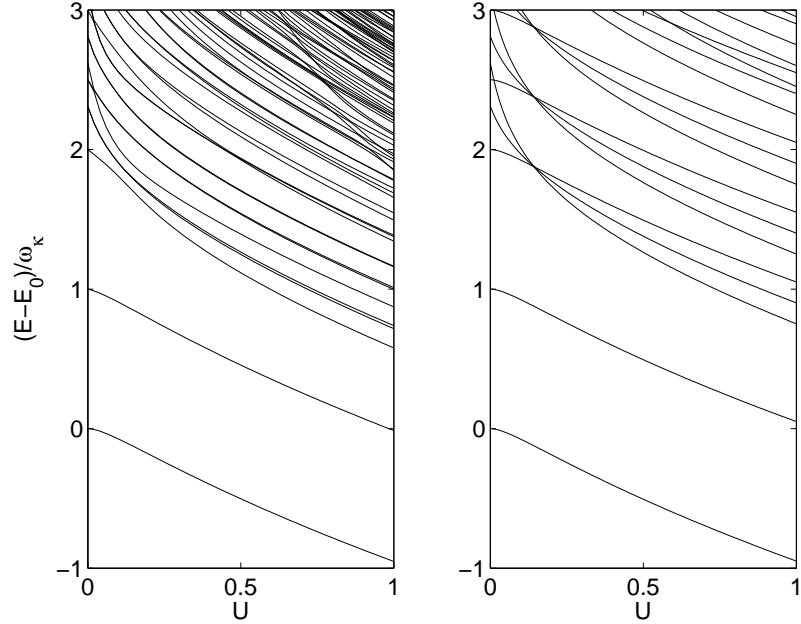}
\caption{Left panel: first few energy levels of $N=25$ atoms in a lattice
with $L=5$ sites, as function of the on-site interaction constant.
(Only the levels corresponding to zero total quasimomentum are shown.)
Right panel: first few energy levels of the Bogoliubov spectrum,
calculated on the basis of Eq.~(\ref{29}) and Eq.~(\ref{29a}).}
\label{fig6}
\end{figure}

It is interesting to compare the discussed Bogoliubov spectrum
of an infinite system with the low-energy spectrum
of a finite system. For this reason we calculate numerically
the spectrum of $N=25$ atoms in a lattice with $L=5$ sites.
\footnote{For $L=5$ there are two different frequencies $\omega_k$.
In this sense, $L=5$ is the simplest generic case to discuss
the Bogoliubov spectrum.}
A few first levels of this system are depicted in the left panel of
Fig.~\ref{fig6} where, to facilitate a comparison, we
substract the energy $E_0=-JN+UN(N-1)/2L$ and rescale energy axis on
the basis of the frequency $\omega_1=\omega_1(U)$.
The Bogoliubov spectrum, calculated by using Eqs.~(\ref{29}-\ref{29a}), 
is depicted in the right panel of Fig.~\ref{fig6}. Both similarities and
differences are evident. The first two levels are seen to
coincide in the whole interval $0\le U\le 1$. On the other hand,
multiple degeneracy of the levels around $U=0.13$ in the right panel
is removed in the left panel. This is, in fact, not surprising.
Indeed, let us consider the lowest group of levels, showing the degeneracy.
These levels are associated with the qusimomentum Fock
states $|0,2,21,2,0\rangle$, $|0,2,22,0,1\rangle$, $|1,0,22,2,0\rangle$,
and $|1,0,23,0,1\rangle$. (Here we use a different notation for the 
Fock states, corresponding to the Brillouin zone $-\pi<\kappa\le\pi$.)
These states belong to different spectra, labeled by $k$ and $m$
in Eq.~(\ref{29}), and are decoupled within the Bogolubov approach. 
However, for the considered finite system these states are coupled by
interaction, where the coupling matrix elements are of the order of $g/N$.

\subsection{Persistent current}
\begin{figure}[t]
\center
\includegraphics[height=8.5cm, clip]{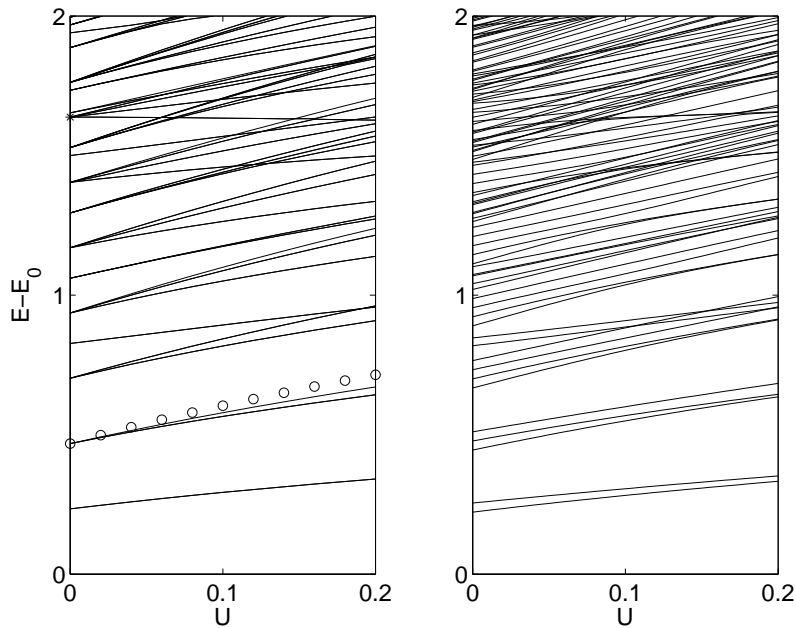}
\caption{Low-energy levels of $N=7$ atoms in a lattice
with $L=9$ sites for magnitude of the scattering potential
$\epsilon=0$ (left) and $\epsilon=0.2$ (right). The state with supercurrent 
($\kappa=2\pi/L$) corresponds to the lowest level in the
group of levels marked by an asterisk. Open circles corresponds
to the quasiparticle energy $\hbar\omega_1$.}
\label{fig7}
\end{figure}

In this subsection we critically review the result of Sec.~\ref{sec2}
about the persistent current, carefully checking validity
of the perturbative approach. Left panel in Fig.~\ref{fig7}
shows the low-energy levels for $N=7$ and $L=9$ (whole spectrum is
shown, i.e., no symmetry selection according to the total quasimomentum).
Remarkably, even for such a small number of atoms one still has a
qualitative agreement with the Bogoliubov spectrum.  Our states
of interest in Fig.~\ref{fig7} are the supercurrent and the
intermidiate states $|\kappa(m)\rangle$, which originate
from the point marked by an asterisk. It is seen, by comparing
with Fig.~\ref{fig7b}(a), that (i) the splitting
between these levels well matches Eq.~(\ref{6}) and (ii) the coupling
of these states to the other states of the system is negligible,
which is indicated by absence of the avoided crossings. 
In addition to the case $\epsilon=0$, right panel
in Fig.~\ref{fig7} shows the spectrum of the atoms in the presence of a weak
scattering potential [should be compared with Fig.~\ref{fig7b}(b)].
Again, no avoided crossings with the other
levels are seen. Hence, the approach of Sec.~2 is well justified.

\begin{figure}[t]
\center
\includegraphics[height=8.5cm, clip]{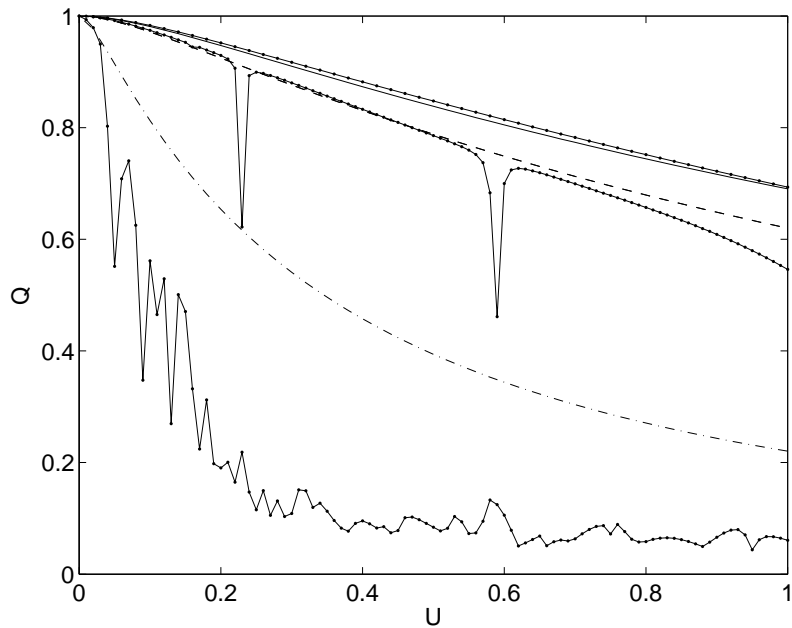}
\caption{Overlap of the states $|\kappa=0\rangle$ (solid line),
$|\kappa=2\pi/L\rangle$ (dashed line), and $|\kappa=4\pi/L\rangle$ (dash-dotted
line) with the Bogoliubov states, originating from these supercurrent states.
Dots (guided by the solid lines) show the quantity (\ref{32}), calculated
for these three values of the initial quasimomentum. The system parameters are
the same as in Fig.~\ref{fig7}(a).}
\label{fig10}
\end{figure}

The above visual analysis of the spectrum can be made
quantitative by considering the overlap of the supercurrent state 
$|\kappa\rangle$ with the exact eigenstates,
\begin{equation}
\label{32}
Q(U)=\max_{n=1}^{\cal N} \left(|\langle\kappa|\Psi_n(U)\rangle|^2\right)\;.
\end{equation}
For $\kappa=0$ the quantity (\ref{32}) is obviously maximazed
by the ground state $|\Psi_0(U)\rangle$, which is expected to
coincide with the Bogoliubov state ground. The solid line in
Fig.~\ref{fig10} shows the overlap of the state $|\kappa=0\rangle$
with the ground Bogoliubov state. A monotonic decrease of $Q=Q(U)$, 
seen in the figure, is due to population of the single-particle 
quasimomentum states with $\kappa\ne0$, and is often
referred to as the Bogoliubov depletion of the BEC.
Additionally, the dashed and dash-dotted lines in Fig.~\ref{fig10}
depict the overlap of the states
$|\kappa=2\pi/L\rangle$ and $|\kappa=4\pi/L\rangle$
with the Bogoliubov states, originating from these supercurrent states,
which we calculate by substituting the single-particle
excitation energy $E_k=J[1-\cos(2\pi k/L)]$ in Eq.~(\ref{23}) by
\begin{eqnarray}
\label{33}
\fl E_k=0.5J[\cos(\kappa+2\pi k/L)+\cos(\kappa-2\pi k/L)-2\cos\kappa]
\sim (2\pi k/L)^2\cos\kappa \;.
\end{eqnarray}
Finally, the series of dots correspond to the
quantity (\ref{32}). It is seen that for $\kappa=0$ dots perfectly
follow the solid continuous line. Thus the ground state of the system is indeed
well approximated by the Bogoliubov state. With exception of
two narrow avoided crossings this is also the case for
the state of our interest $\kappa=2\pi/L$. However, for higher
initial quasimomentum $\kappa=4\pi/L$, the Bogoliubov state is seen
to be completely destroyed by the large number of avoided crossings.

\section{Conclusions}
\label{sec5}
Within the formalism of the Bose-Hubbard model we have considered
time evolution of the atomic supercurrent
in a ring optical lattice with weak on-site disorder.
For vanishing atom-atom interactions, weak disorder induces Rabi
oscillations of the atomic current, where the atoms periodically
change their velocity to the opposite one. For non-vanishing
atom-atom interactions, the supercurrent dynamics
depend crucially on the initial quasimomentum $\kappa$
(i.e., the initial velocity of the atoms). Namely, for a high quasimomentum
$\kappa\sim \pi/2$ the supercurrent exponentially decays as 
the interaction constant $U$ exceeds some critical value,
while for a low quasimomentum $\kappa\ll \pi/2$ the
oscillatory behaviour of the suppercurrent changes to a persistent
current.

The explanation for these effects is found in the structure of
low- and high-energy spectra of the Bose-Hubbard model. It is
shown that the low-energy spectrum of the system is regular,
and the positions of the energy levels can be found by using
a Bogoliubov approach.
In contrast, the high-energy spectrum shows a transition
from regular to a chaotic one if $U$ exceeds its critical value.
Using the results of the random matrix theory, we show
that this transition is reflected in the exponential decay 
of the supercurrent with the decay constant proportional to $U^2$.

This work was supported by Deutsche Forshungsgemeinschaft
within  the SPP1116 program.

\vspace*{0.5cm}

\end{document}